\documentclass[lettersize,journal]{IEEEtran}
\usepackage{amsmath,amsfonts}
\usepackage{algorithmic}
\usepackage{algorithm}
\usepackage{array}
\usepackage[caption=false,font=normalsize,labelfont=sf,textfont=sf]{subfig}
\usepackage{textcomp}
\usepackage{stfloats}
\usepackage{url}
\usepackage{verbatim}
\usepackage{graphicx}
\usepackage{cite}
\usepackage{xcolor}
\usepackage{bm}
\begin{document}

\title{Shunt-controlled resistive state of superconducting wires}

\author{K. Harrabi\textsuperscript{1,2,*}, Z. Alzoubi\textsuperscript{1}, L. R. Cadorim\textsuperscript{3}, and M. V. Milo\v{s}evi\'c\textsuperscript{3,$\dagger$}
\thanks{\textsuperscript{1}Physics Department, King Fahd University of Petroleum and Minerals, 31261 Dhahran, Saudi Arabia}
\thanks{\textsuperscript{2}Interdisciplinary Research Center (RC) for Advanced Quantum Computing, KFUPM, Dhahran 31261, Saudi Arabia}
\thanks{\textsuperscript{3}COMMIT, Department of Physics, University of Antwerp, Antwerp 2000, Belgium}
\thanks{\textsuperscript{*}Email: harrabi@kfupm.edu.sa}
\thanks{\textsuperscript{$\dagger$}Email: milorad.milosevic@uantwerpen.be}}



\maketitle

\begin{abstract}
The use of resistive shunts in superconducting electronics is vast and versatile, to dampen oscillations in junctions, stabilize switching behavior, aid current sensing, divert current during quenches, and protect both the superconductor and the circuit from damage. In single-photon detection by superconducting nanowires, the shunt is crucial for the timely relaxation of the sensor between the events to detect. Here we step out from the superconducting state and discuss the effect of the shunt resistor on the resistive state of a superconducting wire, at elevated currents still below the critical current for the transition to the normal state. We reveal how the shunt resistance controls the system dynamics and the onset of different resistive phases that include hot-spot and phase-slippage events. The accompanying dynamic current redistribution in the circuit also affects the local heating properties and additionally contributes to the control of the resistive state, particularly important at the elevated operation temperatures. 
\end{abstract}


\section{Introduction}
\IEEEPARstart{T}{he} resistive state of superconductors, where dissipation arises from the motion of Abrikosov vortices, plays a central role in a wide range of superconducting technologies. Understanding, and ultimately controlling, the vortex dynamics under different physical conditions is one of the most important challenges in modern days electronics \cite{villegas2003,de2006,savel2002,silhanek2003,milovsevic2011,kalcheim2017,lyu2020,niedzielski2025,zhakina2024}.

A particularly relevant example, where the manipulation of the resistive state becomes possible, is the case of superconducting circuits with shunt resistances. In such systems, diverting part of the applied current from the superconducting wire to the shunt resistor provides a controlled dissipation pathway that allows for the recovery of the superconducting state, preventing the heating associated with the resistive state from stopping the operation of the superconducting element and also protecting the surrounding circuitry.
Resistive shunts have been used for years in applications such as superconducting nanowire single-photon detectors (SNSPDs) \cite{natarajan2012,marsili2013,holzman2019,vodolazov2015} and spiking neuron networks (SNNs) \cite{toomey2018,toomey2019,toomey2020,schneider2022,karamuftuoglu2025,schegolev2023,karimov2024,segall2023,islam2023}.

In this work, we experimentally and numerically investigate how the shunt resistance affects the resistive state of superconducting wires. Discussing how the dynamic current redistribution between the superconductor and different shunt resistances allows the control of the superconducting resistive state and its resistance.
The paper is structured as follows. In Sec.~\ref{sec1}, we introduce our system and the experimental procedures. In Sec.~\ref{sec2} we detail our numerical formalism. The results are presented and discussed in Sec.~\ref{sec3}. Finally, we present our concluding comments in Sec.~\ref{sec4}.

\section{Experimental Setup\label{sec1}}
Thin NbTiN whiskers with a thickness of (\(20\,\mathrm{nm}\) and a width of \(10\,\mu\mathrm{m}\) were fabricated on sapphire and silicon-oxide substrates (STAR Cryoelectronics, New Mexico, USA) under high-vacuum conditions. Superconducting filaments and gold contacts were defined using photolithography followed by ion milling. Figure 1 shows the sample layout, which includes a central narrow stripe with a length of approximately \(\sim 800\,\mu\mathrm{m}\). For biasing and readout, off-chip metallic shunt resistors with resistance values \(R_{\mathrm{sh}} \in \{0.03,\,0.05,\,0.075,\,1.3\}\,\Omega\)
were mounted in parallel with the superconducting wire, with one resistor value used per measurement run. In the measurement configuration, a voltage pulse was applied to one end of the strip, while the opposite end, together with the parallel shunt resistor, was connected to ground. All measurements were performed under vacuum using a closed-cycle cryostat to ensure temperature stability. The sample exhibited a superconducting transition temperature of approximately \(T_c \approx 11\,\mathrm{K}\).
The NbTiN film growth took place in an argon–nitrogen plasma inside a high-vacuum chamber (STAR Cryoelectronics, NM, USA) and was sputtered onto a sapphire substrate. Four gold pads with a thickness of \(100\,\mathrm{nm}\) were patterned using photolithographic and ion-milling processes and were used as electrical contacts.

\begin{figure*}[!t]
\centering
\includegraphics[width=0.9\linewidth]{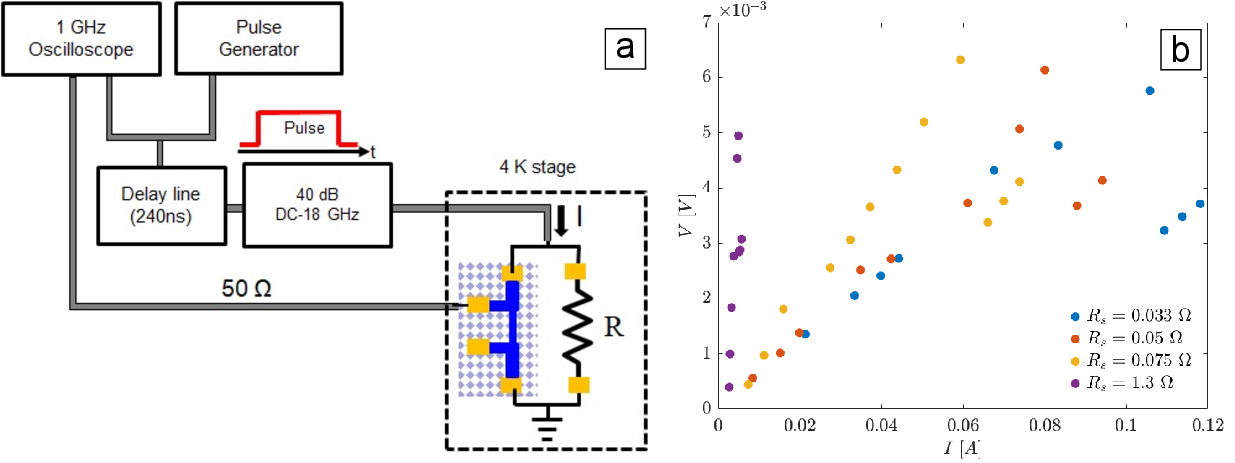}
\caption{\textbf{Experimental setup and measured resistance of a shunted superconducting wire.} (a) Schematic depiction of the sample and the measurement setup, showing a lateral voltage probe for the superconducting wire shunted by a resistor. The sample is current biased, using a pulse generator, a series of calibrated attenuators, and a delay line essential for isolating the incident pulse from the reflected pulse, all connected via 50 $\Omega$ coaxial cables.
 (b) The current-voltage characteristics of a superconducting stripe (width 10$\mu$m, $T=9$K) in the resistive state, for different shunt resistances. Two stages are visible in the resistive state for all shunts used, with a negative differential resistance transition between them.}
\label{fig1}
\end{figure*}

After stabilizing the system at the base temperature \(T = 5.0\,\mathrm{K}\), we applied variable-amplitude voltage pulses along a \(50\,\Omega\) coaxial line through calibrated attenuators. Each pulse had a fixed duration of \(450\,\mathrm{ns}\) and a repetition rate of \(10\,\mathrm{kHz}\), and was routed through a \(240\,\mathrm{ns}\) air-delay line to temporally separate the incident and reflected waves at the device. A high-speed oscilloscope simultaneously monitored the voltages across the superconducting wire and the metallic shunt resistor; the voltage waveform at the device was picked up from a \emph{lateral probe} adjacent to the strip, while the transport current flowed along the superconducting strip. In the superconducting state the sample behaves as an effective short, producing a negative reflection \(\Gamma \approx -1\); accordingly, the device current was obtained from transmission-line relations as \(I = 2V/Z_0\) with \(Z_0 = 50\,\Omega\). For \(I < I_c\) the measured voltage was zero; once the drive exceeded \(I_c\), a finite voltage transient appeared and was recorded. By sweeping the pulse amplitude, we collected a series of current--voltage pairs and compiled the resulting \(I\!-\!V\) characteristics. The full measurement sequence was repeated at multiple stabilized temperatures---ranging from the \(5\,\mathrm{K}\) base up to values near the superconducting transition---using identical pulse timing and delay settings, yielding temperature-dependent \(I\!-\!V\) curves.

\section{Theoretical Formalism\label{sec2}}

To support our experimental results and improve our understanding of the resistive state dynamics, we have numerically modeled our circuit by solving the time-dependent Ginzburg-Landau (TDGL) equation \cite{chapman1996}:
\begin{eqnarray}
    \frac{u}{\sqrt{1+\gamma^2|\psi|^2}}\left ( \frac{\partial }{\partial t}
    +\text{i}\varphi + \frac{\gamma^2}{2}\frac{\partial |\psi|^2}{\partial t^2} \right ) \psi = \nonumber \\  
    \left(\mbox{\boldmath $\nabla$}-i\textbf{A}\right)^2\psi
    +\psi(1-|\psi|^2),
    \label{eq:eq1}
\end{eqnarray}
where we express the order parameter $\psi$ in units of the order parameter at the Meissner state $\psi_{\infty}$; length is presented in units of the superconducting coherence length $\xi$; field and vector potential are expressed in units of $H_{c2}$ and $H_{c2}\xi$, respectively, with $H_{c2}$ being the upper critical field $\Phi_0/2\pi\xi^2$; time in units of $t_{GL}=\pi\hbar/8uk_BT_c$, with $u = 5.79$ being a constant stemming from the microscopic theory. The parameter $\gamma$ is the product of inelastic electron-phonon scattering time and the GL gap at $T = 0$, $\Delta_{GL}$. Without loss of generality, we take $\gamma = 10$ in our calculations. Finally, the scalar potential $\varphi$ is given in units of $V_{GL}\hbar/2et_{GL}$. At each time step, the scalar potential is obtained by solving the Poisson equation:
\begin{equation}
    \bm{\nabla}^2\varphi = \bm{\nabla}\bm{J}_s,
    \label{eq:eq2}
\end{equation}
where the supercurrent is given by:
\begin{equation}
    \bm{J}_s = \textrm{Im}\left [\bar{\psi}{(\mbox{\boldmath $\nabla$}-i\textbf{A})}\psi\right].
\end{equation}
Eqs.~\ref{eq:eq1}-\ref{eq:eq2} are then numerically solved (see Ref.~\cite{milovsevic2010} for details on the method) for a system of size $L_x\times L_y = 100\xi\times100\xi$. The current density $J_s$ transported by the superconductor is introduced through the boundary conditions of the scalar potential as $\bm{\nabla}\varphi(x,\pm L_y/2) = J_s\bm{\hat{y}}$. At the remaining boundaries we have $\bm{\nabla}\varphi(\pm L_x/2,y) = 0$. For the order parameter, we have $\psi(x,\pm L_y/2) = 0$ at the normal contacts and ${(\mbox{\boldmath $\nabla$}-i\textbf{A})}\psi|_x = 0$ at $x = \pm L_x/2$, assuring no supercurrent flows out of the system. The calculations are carried out at zero external magnetic field.

To incorporate the effects of the shunt, we simultaneously solve for the time evolution of the transport current density $J_s$ in a shunted electrical system. This is given by:
\begin{equation}
    L_K\frac{dJ_s}{dt} = (J-J_s)R_s-V_s,
    \label{eq:eq3}
\end{equation}
where $J$ is the total current density of the circuit; $L_K$ is the circuit kinetic inductance presented in units of $L_{GL} = (\Phi_0e^*\pi/8k_bu)/(\sigma_n\xi T_c)$, with $\sigma_n$ being the normal state conductivity of the superconductor material; $R_s$ is the shunt resistance, in units of $R_{GL} = 1/(\sigma_n\xi)$; and $V_s$ is the voltage drop across the superconducting stripe. At each time step, Eq.~\ref{eq:eq3} is solved and the current density carried by the superconductor $J_s$ is recalculated.

\section{Results and Discussion\label{sec3}}

Fig.~\ref{fig1}$(b)$ shows the experimentally measured \(I\!-\!V\) curves at $T = 9$ K and for different values of shunt resistance $R_s$. Here, we focus on the resistive state region, where a finite voltage drop is measured, but the superconducting stripe is not yet in the normal state. As can be seen, for a given value of $R_s$, the voltage increases almost linearly with the current. More importantly, we also observe that, at a fixed current, the measured voltage increases with $R_s$, meaning that the dynamic resistance of the stripe in its resistive state increases with the shunt resistance. Interestingly, we observe that all \(I\!-\!V\) curves present a clear voltage drop at a shunt-dependent critical current, resulting in a negative differential resistance, an important phenomenon previously found in different superconducting systems \cite{reichhardt1997,pedersen1980,ustavcshikov2024}.

\begin{figure}[!t]
\centering
\includegraphics[width=0.9\linewidth]{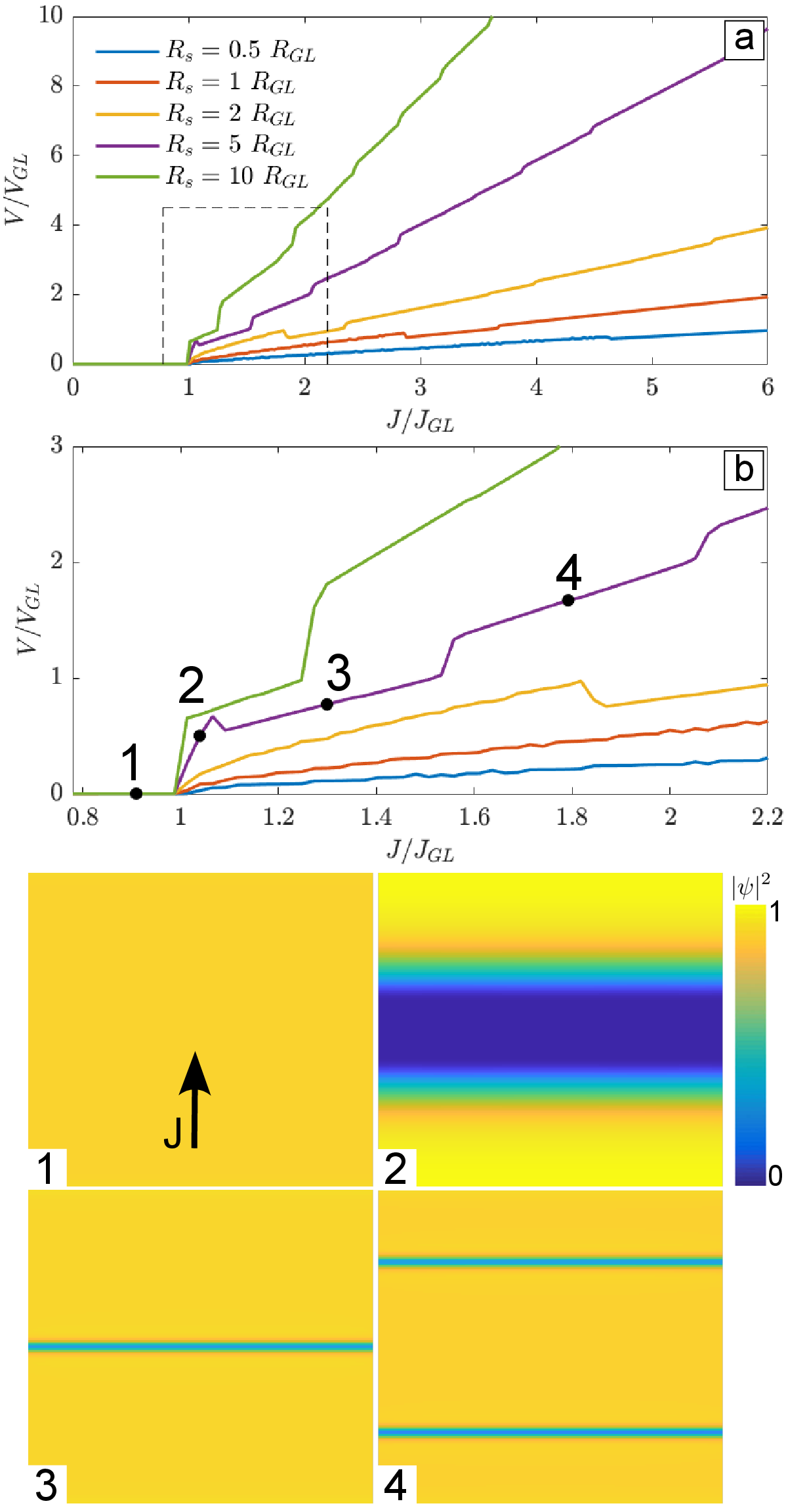}
\caption{\textbf{Simulated I-V curves and visualization of the resistive states.} (a) I-V curves as calculated from TDGL for different values of the shunt resistance. Panel (b) is a zoom on the panel (a), for better visibility of the transitions in the resistive state. The representative snapshots 1-4 of the selected dynamic states are shown below as the plots of the condensate density.}
\label{fig2}
\end{figure}

To understand this behavior, we present in Fig.~\ref{fig2}$(a)$ the numerically calculated \(I\!-\!V\) curves for different values of $R_s$. One can see that both characteristic features of the experimentally measured \(I\!-\!V\) characteristics, namely the nearly linear voltage dependence on the current and the increase of dynamical resistance (\textit{i.e.} voltage) with $R_s$, are reproduced by the TDGL model. 

A more detailed view of the numerical \(I\!-\!V\) curves is presented in Fig.~\ref{fig2}$(b)$, where we show the \(I\!-\!V\) region highlighted by the dashed black lines in panel $(a)$. As exemplified by the green \(I\!-\!V\) curve, for $R_s = 5~R_{GL}$, current regions with different dynamic resistances can be identified. To understand the dynamics underlying these different resistive states, we show snapshots of the Cooper-pair density in the resistive state for four current values. State $1$ represents the range of small currents, below the Ginzburg-Landau depairing current $J_{GL}$, where the external current is not sufficiently strong to break the Cooper-pairs and, as a result, the system stays within the dissipationless Meissner state, characterized by the homogeneous order-parameter shown in inset $1$.

This homogeneous state is broken when the current through the superconductor reaches $J_{GL}$. At this point, the superconductor can no longer support current transport without dissipation and the resistive state sets in. The dynamic behavior of the resistive state depends on both the applied current value and the shunt resistance, as detailed below. Initially, for all values of $R_s$ (except $R_s = 10\Omega$) the system displays a state with a large dynamical resistance, where the superconducting state is initially suppressed along the entire sample. Due to the large emergent resistance of the normal state, the current is then diverted to the shunt, allowing superconductivity to fully recover. Inset $2$ shows a time instant where the normal region (dark blue) decreases while the superconducting state is recovered.
This process is then periodically repeated, in a dynamical equilibrium.

To understand why this state is absent for $R_s = 10\Omega$, we need to first discuss what happens when we continue to increase the applied current and the system transits to a new behavior. As the current increases, the frequency of the above described process of destruction/recovery of the superconducting state also increases. Eventually, the system reaches a point where the Meissner state is no longer recovered and, instead, a line of suppressed superconductivity (a quasi-phase-slip line (PSL) \cite{vodolazov2007,berdiyorov2009}) is stabilized. Such a state is depicted in inset $3$. As can be seen from panels $a)$ and $b)$, the current at which the transition from states $2$ and $3$ occurs decreases with increasing $R_s$. This is due to the fact that the frequency of the dynamical states increases with $R_s$ (see the more detailed discussion below). As a consequence, the required current for the emergence of the PSL decreases and, for $R_s = 10\Omega$, it is equal to $J_{GL}$, resulting in the absence of state $2$.

It is important to note that the width of PSL does not increase \textit{i.e.} the superconductor no longer goes fully to the normal state. As a result, the dynamical resistance is much smaller than the one found for state $2$. In fact, one can see that a voltage drop occurs at the passage from state $2$ to state $3$, causing the negative differential resistance as observed in the experimental data of Fig.~\ref{fig1}. As the total applied current is increased even further, additional quasi-PSLs are formed along the stripe, as shown in state $4$, leading to additional voltage jumps in the \(I\!-\!V\) characteristics. We note that in real experimental conditions, the heating associated with the normal current of the quasi-PSLs suppresses these voltage jumps, favoring a smoother \(I\!-\!V\) characteristic in the phase-slip regime.

\begin{figure*}[!t]
\centering
\includegraphics[width=\linewidth]{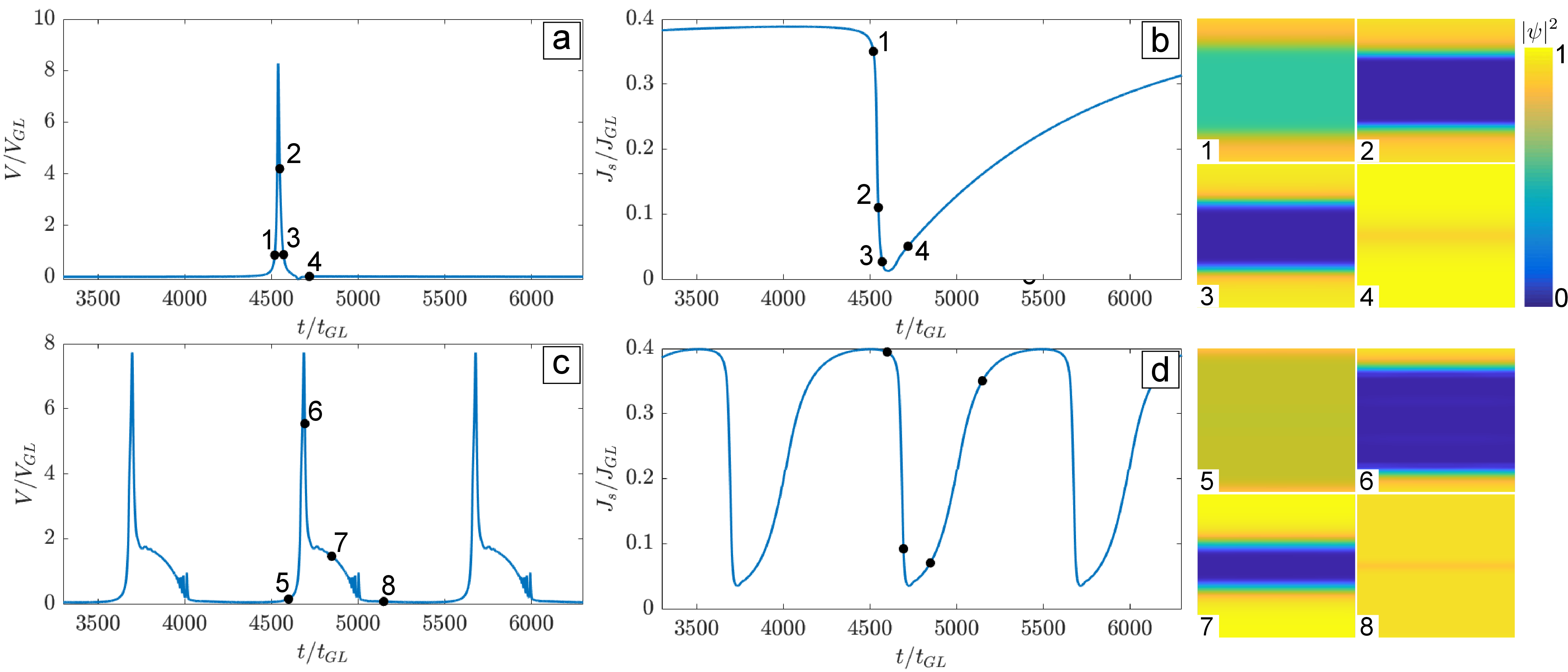}
\caption{\textbf{Temporal characterization of the resistive states}. Voltage and supercurrent versus time, corresponding to state $2$ from Fig.~\ref{fig2} (applied current density $J=1.06J_{GL}$), for two values of the shunt resistance: $R_s=0.5R_{GL}$ (a,b) and $R_s=10R_{GL}$ (c,d). The observed temporal features and characteristic times are in direct correlation with condensate dynamics shown through selected snapshots on the right.}
\label{fig3}
\end{figure*}

Having discussed the voltage dependence on applied current, we proceed with the investigation of how the resistive dynamics is affected by the shunt. Fig.~\ref{fig3}$(a)$ shows the simulated voltage as a function of time for $R_s = 0.5~R_{GL}$ and $J = 1.06~J_{GL}$. As can be seen, the signal consists of a single voltage peak corresponding to the process of destruction/recovery of the superconducting state that was detailed above. The dynamics of such a process is shown in insets $1-4$, where we plot the Cooper-pair density at different instances of time.

To better understand the contribution of the shunt to this dynamics, we plot in panel $(b)$ the current density flowing in the superconductor, $J_s$, as a function of time. When the first peak of voltage takes place in the superconducting stripe due to the onset of a normal region, the electric current is diverted from the superconductor to the shunt. During this process, the supercurrent density $J_s$ decreases exponentially with a characteristic time $\tau = L/(R_s+R_w)$ \cite{toomey2019}, where $L$ is the kinetic inductance of the circuit and $R_w$ is the resistance of the superconductor in the resistive state. As expected, $R_w$ varies with time and depends on the dynamics of the normal region in the stripe. For example, when superconductivity starts to get suppressed, $R_w$ abruptly increases and we observe a pronounced dip in $J_s$. When $R_w$ becomes sufficiently small, the current diverted to the shunt returns to the superconductor. In this stage, $J_s$ exponentially increases on the timescale of $\tau$. Since now $R_w$ is smaller than before, current returns to the superconductor at a slower rate. This can be clearly seen by comparing the slopes of $J_s(t)$ before and after state $3$ in panel $(b)$. This entire process is repeated at a later time, not shown in the figure.

Let us now compare the dynamics of the resistive state at the same current density $J = 1.06~J_{GL}$ for a larger shunt, $R_s = 5~R_{GL}$. The voltage and supercurrent density as functions of time for this case are shown in panels $(c)$ and $(d)$ of Fig.~\ref{fig3}, respectively. As illustrated by states $5-8$, the dynamics of the normal state region is very similar to that discussed for $R_s = 0.5~R_{GL}$, exhibiting the same sequence of events. 

The main difference between the systems with two different shunts is the timescale over which this process occurs, which decreases as the shunt resistance increases. This can be readily seen by comparing the number of voltage peaks in panels $(a)$ and $(c)$. Quantitatively, the time interval between the largest voltage peaks, represented by states $2$ for $R_s = 0.5~R_{GL}$ and $6$ for $R_s = 10~R_{GL}$, is $\Delta t = 4929~\tau_{GL}$ and $\Delta t = 991~\tau_{GL}$, respectively. The reduction in $\Delta t$ arises from the shorter characteristic time $\tau = 1/(R_s+R_w)$ required for the current to redistribute between the shunt and the superconductor, as can be seen by comparing the $J_s(t)$ curves in panels $(b)$ and $(d)$. Notice however that the 10-fold increase in $R_s$, from $0.5$ to $10~R_{GL}$, is not reflected in a 10-fold decrease in $\Delta t$. This occurs because the process of annihilation and nucleation of the superconducting state has its own characteristic timescale, which limits the variation of the $V(t)$ period with varied shunt resistance. At larger applied currents, the difference between the systems with different shunts becomes even more dramatic, as the shorter characteristic time $\tau$ for larger shunt resistance favors the emergence of quasi-PSL's at smaller currents, leading to the voltage jumps and pronounced resistance difference observed in the \(I\!-\!V\) characteristics of Fig.~\ref{fig2}.

\section{Conclusion\label{sec4}}
In summary, we have shown how shunt resistances can be used to control the resistive response of a superconducting wire. Our experimental and numerical results have shown that two different resistive states (hot spot or phase slip) can be stabilized in a superconducting stripe, both of which exhibit increasing resistance with increasing shunt resistance. Furthermore, the voltage drop at the transition current between the two resistive states decreases as the shunt resistance is made smaller, and requires a larger critical current as the shunt resistance is made smaller.

Through the numerical solution of the TDGL equations, we have shown the mechanism by which the shunt resistance affects the dynamics of the resistive state in the superconductor, slowing down the annihilation of superconductivity due to the current deviation from the superconductor to the resistor.

In conclusion, our results revealed that the resistive state of superconducting wires can be efficiently controlled using shunt resistances, presenting a pathway to sustain superconductivity at large current densities, and a particularly broad tunability of the superconducting state - useful in prospective sensing and electronic devices based on superconducting stripes.

\section*{Acknowledgments}
The authors gratefully acknowledge the support of the Interdisciplinary Research Center (RC) for Interdisciplinary Research Center for Advanced Quantum Computing, King Fahd University of Petroleum and Minerals, under the research grant $INSS2504$.

\bibliographystyle{IEEEtran}
\bibliography{bibliography}

\newpage

\end{document}